\documentclass[journal]{IEEEtran}
\usepackage{graphicx}
\usepackage{epstopdf}
\usepackage{fancyhdr}
\usepackage{amsmath,amssymb,amstext}
\usepackage{subeqnarray}
\usepackage{cite}
\usepackage{hhline}
\usepackage{bm}
\usepackage{soul}
\usepackage{slashbox}
\usepackage{times,amsmath,amsfonts}
\usepackage{subfigure}
\usepackage{amsmath,amssymb}
\usepackage{graphicx,picture}
\usepackage{cite}
\usepackage{subfigure}
\usepackage{subeqnarray}
\usepackage{cases}
\usepackage{color}
\usepackage{overpic}
\usepackage{multirow}
\usepackage{booktabs}
\usepackage{hyperref}
\usepackage{algorithmic}
\usepackage[ruled, linesnumbered]{algorithm2e}
\ifCLASSINFOpdf
\else
   \usepackage{graphicx}
   \graphicspath{{../eps/}}
   \DeclareGraphicsExtensions{.eps}
\fi
\usepackage{fancyhdr}
\usepackage{amsmath,amssymb,amstext}
\usepackage{subeqnarray}
\usepackage{cite}
\usepackage{hhline}
\usepackage{bm}
\usepackage{soul}
\usepackage{slashbox}
\usepackage{times,amsmath,amsfonts}
\usepackage{subfigure}
\usepackage{amsmath,amssymb}
\usepackage{graphicx,picture}
\usepackage{cite}
\usepackage{subfigure}
\usepackage{subeqnarray}
\usepackage{cases}
\usepackage{color}
\usepackage{overpic}
\usepackage{multirow}
\usepackage{booktabs}
\usepackage{epstopdf}

\definecolor{r}{rgb}{1,0,0}
\definecolor{b}{rgb}{0,0,1}
\definecolor{k}{rgb}{0,1,1}
\usepackage{upgreek}

\newcounter{saveeqn}%

\allowdisplaybreaks
\DeclareMathSymbol{\Phi}{\mathord}{letters}{8}

\begin{document}
\title{Achieving Full-Bandwidth Sensing Performance with Partial Bandwidth Allocation for ISAC}

\author{
\IEEEauthorblockN{Zhiqiang Xiao~\IEEEmembership{Member, IEEE}, Zhiwen Zhou, Qianglong Dai, Yong Zeng,~\IEEEmembership{Senior Member, IEEE},\\
 Fei Yang, and Yan Chen,~\IEEEmembership{Senior Member, IEEE}}

\thanks{
Z. Xiao, Z. Zhou, Q. Dai, and  Y. Zeng are with the National Mobile Communications Research Laboratory, Southeast University, Nanjing, 210096, China; Z. Xiao and Y. Zeng are also with the Purple Mountain Laboratories, Nanjing, 211111, China (e-mail: \{zhiqiang\_xiao, yong\_zeng\}@seu.edu.cn). (\emph{Corresponding author: Yong Zeng.})

F. Yang and Y. Chen are with the Wireless Technology Lab., Huawei Technologies Co., Ltd, Shanghai 201206, China.
}
}
\maketitle

\begin{abstract}
This letter studies an uplink integrated sensing and communication (ISAC) system using discrete Fourier transform spread orthogonal frequency division multiplexing (DFT-s-OFDM) transmission.
We try to answer the following fundamental question: With only a fractional bandwidth allocated to the user with sensing task, can the same delay resolution and unambiguous range be achieved as if all bandwidth were allocated to it?
We affirmatively answer the question  by proposing a novel two-stage delay estimation (TSDE) method
that exploits the following facts: without increasing the allocated bandwidth, higher delay resolution can be achieved via distributed subcarrier allocation compared to its collocated counterpart, while there is a trade-off between delay resolution and unambiguous range by varying the decimation factor of subcarriers.
Therefore, the key idea of the proposed TSDE method is to first perform coarse delay estimation with collocated subcarriers to achieve a large unambiguous range, and then use distributed subcarriers with optimized decimation factor to enhance delay resolution while avoiding delay ambiguity.
Our analysis shows that the proposed TSDE method can achieve the full-bandwidth delay resolution and unambiguous range, by using only at most half of the full bandwidth, provided that the channel delay spread is less than half of the unambiguous range.
Numerical results show the superiority of the proposed method over the conventional method with collocated subcarriers.
\end{abstract}


\section{Introduction}
Integrated sensing and communication (ISAC) has been identified as one of the main usage scenarios of the sixth-generation (6G) mobile communication networks \cite{ITU-R}, enabling to exploit the cellular network infrastructure for constructing a ubiquitous perceptive networks.
To this end, extensive research efforts have been devoted to studying various aspects of ISAC\cite{xiao2022waveform,xiao2024simultaneous,zhang2023integrated,pan2023cooperative}, such as waveform design \cite{xiao2022waveform}, beam codebook design \cite{xiao2024simultaneous}, and prototyping experiments \cite{zhang2023integrated}, etc.

For uplink wireless communications, discrete Fourier transform (DFT) spread orthogonal frequency division multiplexing (OFDM), i.e., DFT-s-OFDM \cite{abd2014sc}, is commonly used in the fourth-generation (4G) and fifth-generation (5G) mobile communication systems, due to its low peak-to-average power ratio (PAPR), easy for implementation, and flexibility for time-frequency (TF) resource allocation.
On the other hand, for wireless sensing, the delay and Doppler estimation of DFT-s-OFDM can be decoupled and efficiently processed via applying inverse DFT (IDFT) and DFT operations, respectively.
Therefore, DFT-s-OFDM is regarded as one of the candidate waveforms for uplink ISAC \cite{csahin2020dft,wei2023integrated,wu2023dft}.
However, its delay and Doppler resolutions are constrained by the bandwidth and time duration of the received signals.
In particular, for uplink multi-user ISAC systems, only a fractional bandwidth can be allocated to the user with sensing task.
Thus, the delay resolution is severely limited.
Although some subspace-based super-resolution algorithms can improve the sensing resolution, they require higher computational complexity \cite{zhang2023integrated,schmidt1986multiple,roy1989esprit}.
Some related works were reported to design the OFDM signal pattern using non-continuous TF resources for ISAC \cite{ozkaptan2018ofdm,zhang2024ofdm}.
However, there exists a trade-off between sensing resolution and unambiguous sensing range.
Therefore, the following fundamental question is raised : With only a fractional bandwidth allocated to the user with sensing task, can the same delay resolution and unambiguous range be achieved as if all bandwidth were allocated to it?

In this letter, we will show that the answer to the above question is affirmative.
Specifically, we consider an uplink DFT-s-OFDM based ISAC system, where a base station (BS) aims to communicate with multiple user equipments (UEs) while providing sensing service for the UE with sensing task.
In this case, each user is only allocated with a portion of the bandwidth.
Intuitively, with a contiguous bandwidth allocated to the UE with sensing task, the delay resolution is limited by its allocated bandwidth.
However, if the same bandwidth with decimated subcarriers is allocated to it, higher delay resolution can be achieved without increasing the allocated bandwidth, but at the cost of a reduced unambiguous range.

To exploit this fact, we first derive the trade-off between delay resolution and unambiguous range by varying the decimation factor of subcarriers.
Based on it, we propose a novel two-stage delay estimation (TSDE) method.
The key idea is that the UE with sensing task first transmits uplink signals with contiguous subcarriers for coarse delay estimation, which renders a large unambiguous range.
Based on the coarse estimation results, the decimation factor of subcarriers is optimized and then the UE transmits signals with distributed subcarriers using the optimized decimation factor.
Thus, the higher delay resolution can be achieved to refine the estimation results while avoiding delay ambiguity.
Our analysis shows that the proposed TSDE method can achieve the same delay resolution and unambiguous range of the full-bandwidth, by using only at most half of the full bandwidth, as long as the channel delay spread is less than half of the unambiguous range.
In the most favorite scenario,  by using only a partial bandwidth with $\sqrt{K}$ subcarriers, the delay resolution and unambiguous range of the full-bandwidth with $K$ subcarriers can be achieved, if the channel delay spread is less than the delay resolution of the collocated $\sqrt{K}$ subcarriers.
Therefore, the proposed method is particularly suitable for sensing in the dense scatterer environment with limited bandwidth.

\section{System Model}
\begin{figure} 
  \centering
  \includegraphics[width=0.35\textwidth]{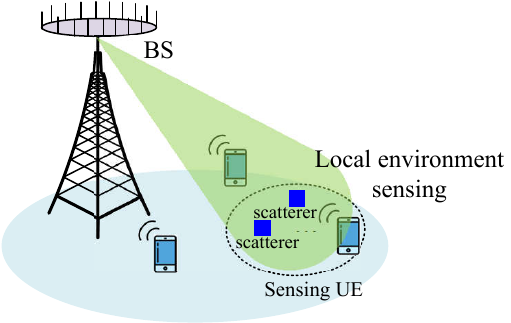}
  \caption{An illustration of the uplink DFT-s-OFDM based ISAC system.}\label{system}\vspace{-0.3cm}
\end{figure}
As illustrated in Fig.~\ref{system}, we consider an uplink ISAC system with DFT-s-OFDM transmission, where the BS wishes to serve multiple UEs for uplink communication while providing local environment sensing for the UE with sensing task via its received multi-path signals.
The system total bandwidth is $B$.
There are totally $K$ subcarriers with the subcarrier spacing $\triangle f = \frac{B}{K}$.
The uplink channel between the sensing UE and the BS at the $k$th subcarrier can be expressed as
\begin{equation}\label{channel}
H_{k} = \sum\nolimits_{l=1}^L{\alpha}_le^{-j2\pi\triangle fk\tau_l}, k=0,\cdots,K-1,
\end{equation}
where $L$ is the number of multi-paths, $\alpha_l$ denotes the $l$th path gain after the receive beamforming, and $\tau_l$ is the propagation delay of the $l$th path.\footnote{
Note that we assume that the receive beam is prior known, say via beam training.
Moreover, as the delay and Doppler estimation for DFT-s-OFDM can be decoupled and processed similarly, we focus on the delay estimation. We comment that the proposed method and analysis results can be directly applied to the Doppler counterparts.
Besides, the proposed method can be also applied to downlink OFDM access (OFDMA) systems.}
Denote by $\tau_d\triangleq\tau_{\max}-\tau_{\min}$ the delay spread for the channel of the sensing UE, where $\tau_{\max}=\max\limits_{1\le l\le L}\tau_l$ and $\tau_{\min}=\min\limits_{1\le l\le L}\tau_l$ are the maximum and minimum propagation delays, respectively.

For uplink transmission, each UE is only assigned a subset of subcarriers.
Let $K_1$ denote the number of subcarriers allocated to the UE with sensing task, with $K_1<K$.
Thus, its allocated bandwidth is $B_1=\triangle fK_1<B$.
For DFT-s-OFDM transmission, each group of $K_1$ information-bearing symbols $x[n]$, $n=0,\cdots,K_1-1$, are processed via $K_1$-point DFT as
\begin{equation}\label{DFT_spread}
X[k] = \sum\nolimits_{n=0}^{K_1-1}x[n]e^{-\frac{j2\pi nk}{K_1}}, k=0,\cdots,K_1-1.
\end{equation}
Next, the outputs in \eqref{DFT_spread} are mapped to $K_1$ out of $K$ subcarriers via subcarrier mapping as
\begin{equation}\label{subcarrier mapping}
\tilde{X}[k] = \left\{
\begin{aligned}
&X[k_1], &&k=\eta k_1 \ \text{with}\ k_1=0,\cdots,K_1-1,\\
&0, &&\text{otherwise},
\end{aligned}
\right.
\end{equation}
where $\eta$ denotes the decimation factor for selecting the subcarriers for the UE with sensing task, with $\eta=1,\cdots,\eta_{\max}$, where $\eta_{\max}\triangleq \frac{K}{K_1}$ is assumed to be an integer for ease of exposition.
There are two common subcarrier mapping schemes, namely {\it localized mapping} for $\eta=1$ and {\it distributed mapping} for $\eta>1$, which leads to {\it collocated subcarriers} and {\it distributed subcarriers}, respectively \cite{abd2014sc}.
To guarantee the sensing performance, we assume that the UE with sensing task has the highest priority to select subcarriers.
After that, \eqref{subcarrier mapping} is converted to the time domain via $K$-point IDFT as
\begin{equation}\label{tx}
\tilde{x}[n] = \frac{1}{{K}}\sum\nolimits_{k=0}^{K-1}\tilde{X}[k]e^{\frac{j2\pi kn}{K}}, n=0,\cdots,K-1.
\end{equation}
Let $T_{\mathrm{cp}}$ be the CP length, with $T_{\mathrm{cp}}=N_{\mathrm{cp}}T_s$ and $T_s=~\frac{1}{B}$.
To avoid the inter-symbol interference (ISI), we assume that  $T_{\mathrm{cp}}\ge \tau_{\max}$.
Thus, the transmit signal including CP is
$\tilde{x}[(n-N_{\mathrm{CP}})\ \mathrm{mod}\ K], n=0,\cdots,K+N_{\mathrm{cp}}-1$.

\section{DFT-s-OFDM based ISAC}
\subsection{Signal Processing}
With the channel in \eqref{channel}, after CP removal, the received signal at the BS from the sensing UE without considering the noise is
\begin{equation}\label{rx1}
\begin{aligned}
\tilde{Y}[k] &= H_k\tilde{X}[k]\\
&=\sum\nolimits_{l=1}^{L}{\alpha}_l\tilde{X}[k]e^{-j2\pi\triangle fk\tau_l}, k=0,\cdots,K-1.
\end{aligned}
\end{equation}
Followed that, the subcarrier demapping is performed, where the frequency-domain samples $\tilde{Y}[k]$ at $k=\eta k_1$ with $k_1=0,\cdots,K_1-1$, are selected, yielding
\begin{equation}\label{output}
Y[k]=\sum\nolimits_{l=1}^L\alpha_lX[k]e^{-j2\pi\triangle f\eta k\tau_l}, k=0,\cdots,K_1-1.
\end{equation}
For communications, after frequency domain equalization (FDE) and $K_1$-point IDFT of \eqref{output}, the information-bearing symbols can be detected.

\begin{algorithm}[t]
	\caption{Successive Delay Estimation}
	\label{alg1}
	\textbf{Input}: $\mathbf{r}\in\mathbb{C}^{K_1\times1}$, $\eta$

    \textbf{Output}: $\hat{\tau}_1,\cdots,\hat{\tau}_{\hat{L}}$, $\mathcal{I}=\{p_1,\cdots,p_{\hat{L}}\}$, $\hat{L}$

    Initialize $\gamma_{\mathrm{th}}$, $\mathcal{I}=\emptyset$, $\gamma = 1$, $\varphi=\|\mathbf{r}\|^2$, $\hat{L}=0$, $\mathbf{F}_{\eta}=\left[\mathbf{f}_0(\eta),\cdots,\mathbf{f}_{N_{\mathrm{cp}}-1}(\eta)\right]\in\mathbb{C}^{K_1\times P}$, where $\mathbf{f}_p(\eta)$, $p=0,\cdots,N_{\mathrm{cp}}-1$, is given by \eqref{f_p}.

    \Repeat{$\gamma \le \gamma_{th}$}{
        $p_l = \arg\max\ |\mathbf{F}_{\eta}[:,p]^H\mathbf{r}|$, $p=0,\cdots,N_{\mathrm{cp}}-1$.

        $\mathcal{I}=\mathcal{I}\cup p_l$, $\hat{L} = \hat{L} + 1$.

        $\mathbf{r} = \mathbf{r}- \mathbf{F}_{\eta}[:,p_l]\left(\mathbf{F}_{\eta}[:,p_l]^H\mathbf{F}_{\eta}[:,p_l]\right)^{-1}\mathbf{F}_{\eta}[:,p_l]^H\mathbf{r}$.

        $\gamma = \|\mathbf{r}\|^2/\varphi$; // update the residual power ratio.
    }
    $\hat{\tau}_l=p_lT_s$, $p_l\in\mathcal{I}$.

\end{algorithm}
On the other hand, for sensing, with the detected information-bearing symbols, the randomness of \eqref{output} can be removed by taking the element-wise division as
\begin{equation}\label{rx}
r[k]=\frac{Y[k]}{{X}[k]} =  \sum\nolimits_{l=1}^L\alpha_le^{-j2\pi\triangle f\eta k\tau_l}, k=0,\cdots,K_1-1.
\end{equation}
Denote by $\mathbf{r}=\left[r[0],\cdots,r[K_1-1]\right]^T\in\mathbb{C}^{K_1\times1}$.
Here, we consider a matched filtering (MF) based method to estimate the multi-path delays.
By sampling the delay axis with $\tau_p=pT_s$ for $0\le \tau_p<T_{\mathrm{cp}}$, the matched filters can be constructed as
\begin{equation}\label{f_p}
\mathbf{f}_p(\eta)=[1,e^{-j2\pi \frac{p\eta}{K}},\cdots,e^{-j2\pi \frac{(K_1-1)p\eta}{K}}]^{T}\in\mathbb{C}^{K_1\times1},
\end{equation}
where $p=0,\cdots,N_{\mathrm{cp}}-1$.
The results after MF are
\begin{equation}\label{est}
\Gamma(\tau_p;\eta)=\mathbf{f}_p^H(\eta)\mathbf{r}/\|\mathbf{f}_p(\eta)\|^2, p=0,\cdots,N_{\mathrm{cp}}-1.
\end{equation}
Therefore, the multi-path delays can be estimated by searching the peaks of $\left|\Gamma(\tau_p;\eta)\right|$, as summarized in Algorithm 1.
Note that
the delay index can be iteratively estimated in {\it Line 4$\sim$9} of Algorithm 1, where
$\mathbf{F}_{\eta}[:,p]$ denotes the $p$th column of $\mathbf{F}_{\eta}$, with $\mathbf{F}_{\eta}=\left[\mathbf{f}_0(\eta),\cdots,\mathbf{f}_{N_{\mathrm{cp}}-1}(\eta)\right]\in\mathbb{C}^{K_1\times P}$ being the sensing matrix.
In {\it Line 5}, the delay index associated with the maximum output of MF is obtained, after which its corresponding signal component is subtracted from $\mathbf{r}$ in {\it Line~7}, to avoid interfering the estimation of other paths.
The loop terminates when the residual signal power ratio $\gamma$ is no greater than a given threshold $\gamma_{th}$.
By doing so, the multi-path delays can be estimated as $\hat{\tau}_l=p_lT_s$, with $p_l\in\mathcal{I}=\{p_1,\cdots,p_{\hat{L}}\}$.

\subsection{Trade-off between Delay Ambiguity and Resolution}
By substituting \eqref{rx} and \eqref{f_p} into \eqref{est}, we have
\begin{equation}\label{gamma}
\begin{aligned}
\Gamma(\tau_p;\eta)&\overset{(a)}{=}\frac{1}{K_1}\sum\nolimits_{l=1}^{L}\alpha_l\sum\nolimits_{k=0}^{K_1-1}e^{-j2\pi\triangle f k\eta(\tau_l-\tau_p)}\\
&=\sum\nolimits_{l=1}^L\alpha_l\mathcal{G}(\tau_p,\tau_l;\eta),
\end{aligned}
\end{equation}
where $(a)$ holds due to $\tau_p=pT_s$, $\triangle f=\frac{B}{K}$, and $T_s=\frac{1}{B}$, and $\mathcal{G}(\tau_p,\tau_l;\eta)$ is the Dirichlet fucntion, which is given by
\begin{equation}
\mathcal{G}(\tau_p,\tau_l;\eta)\triangleq e^{-j2\pi \triangle f(K_1-1)\eta(\tau_l-\tau_p)}\frac{\sin(\pi \triangle fK_1\eta(\tau_l-\tau_p))}{K_1\sin(\pi \triangle f\eta(\tau_l-\tau_p))}.
\end{equation}
It is observed from \eqref{gamma} that the delay sensing performance is determined by $|\mathcal{G}(\tau_p,\tau_l;\eta)|$, which is further affected by $\eta$.

\begin{figure} 
  \centering
  \includegraphics[width=0.45\textwidth]{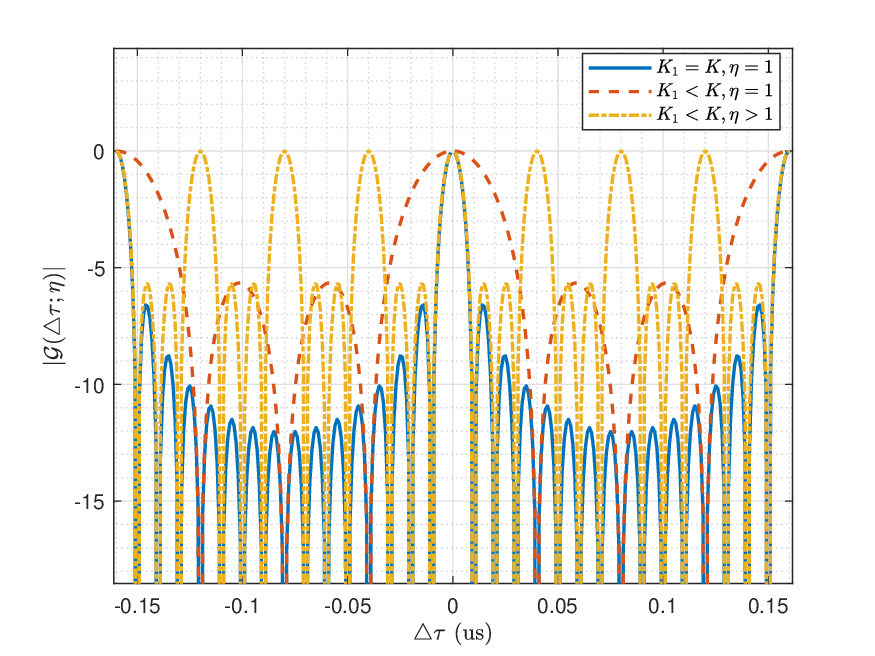}
  \caption{The delay resolution and unambiguious range versus $K_1$ and $\eta$.}\label{delay_range}\vspace{-0.3cm}
\end{figure}
Denote by $\triangle\tau\triangleq\tau_l-\tau_p$, thus $|\mathcal{G}(\tau_p,\tau_l;\eta)|$ can be rewritten as $|\mathcal{G}(\triangle\tau;\eta)|=\frac{\sin(\pi \triangle fK_1\eta\triangle\tau)}{K_1\sin(\pi \triangle f\eta\triangle\tau)}$.
Specifically, as illustrated in Fig.~\ref{delay_range}, the delay resolution is usually defined as the half width of the main lobe of $|\mathcal{G}(\triangle\tau;\eta)|$, which can be obtained by letting $|\pi\triangle fK_1\eta\triangle\tau|=\pi$.
Thus, the delay resolution is
\begin{equation}\label{res}
\tau_{\mathrm{res}}(\eta)= \frac{1}{\triangle fK_1\eta}.
\end{equation}
On the other hand, the grating lobes of $|\mathcal{G}(\triangle\tau;\eta)|$ will appear at $\triangle \tau=\frac{\kappa}{\triangle f \eta}$, $\kappa=\pm1,\cdots,\pm\eta$, which may introduce ambiguity in delay sensing.
Thus, the unambiguous delay range is defined as
\begin{equation}\label{uamb}
\tau_{\mathrm{u}}(\eta) = \frac{1}{\triangle f \eta}.
\end{equation}

\begin{algorithm}[t]
	\caption{Proposed Two-Stage Delay Estimation (TSDE)}
	\label{alg1}
	\textbf{Input}: $K$, $K_1$

    \textbf{Output}: $\hat{\tau}_l$, $l=1,\cdots,\hat{L}$.
	
    // \textbf{Stage-1: Coarse Delay estimation}

    UE transmits uplink signals with collocated subcarriers ($\eta=1$) according to \eqref{DFT_spread}-\eqref{tx}.

    BS processes received signals for sensing according to \eqref{rx1}-\eqref{rx}, outputs $\mathbf{r}=[r[0],\cdots,r[K_1-1]]^{T}\in\mathbb{C}^{K_1\times1}$.

    Perform delay estimation based on Algorithm 1, and output the estimated delay index set $\mathcal{I}$ and the number of estimated multi-paths $\hat{L}$.

    Obtain the delay bin set $\mathcal{U}=\mathrm{unique}(\{\lfloor \frac{p_l}{\eta_{\max}}\rfloor,p_l\in\mathcal{I}\})$, with $\mathcal{U}=\{u_1,\cdots,u_{L'}\}$ and $L'\le \hat{L}$.

    Define searching region, $\Omega \triangleq \Omega_1\cup\cdots\cup\Omega_{L'}$ with $\Omega_{l}\triangleq \left[u_l\eta_{\max},(u_l+1)\eta_{\max}\right)$, $l=1,\cdots,L'$.

    Find the subcarrier decimation factor $\eta^*$. (Section IV-B)

    // \textbf{Stage-2: Refined Delay estimation}

    Repeat Step 4-6 with $\eta=\eta^*$, refining the estimation results in the region $\Omega$ based on Algorithm 1, and output estimated multi-path delays as $\hat{\tau}_l$, $l=1,\cdots,\hat{L}$.

\end{algorithm}
{\it Remark 1}:
There is a trade-off between delay resolution $\tau_{\mathrm{res}}(\eta)$ and unambiguous range $\tau_{\mathrm{u}}(\eta)$ by varying $\eta$.
As illustrated in Fig.~\ref{delay_range}, when $\eta=1$, the unambiguous range is large but the resolution is low.
Without increasing the total allocated bandwidth $B_1=\triangle f K_1$, by increasing $\eta$, higher resolution can be achieved but the unambiguous range is reduced.
In particular, when $\eta=\eta_{\max}=\frac{K}{K_1}$, the full-bandwidth resolution $\frac{1}{B}$ can be achieved but with the smallest unambiguous range $\frac{K_1}{B}$ according to \eqref{res} and \eqref{uamb}.

In the following, by leveraging the trade-off in {\it Remark 1}, we propose a novel TSDE scheme.

\section{Proposed Two-Stage Delay Estimation Scheme}
The proposed method consists of two stages, i.e., coarse delay estimation and refined delay estimation.
The key idea is that the sensing UE first transmits uplink signals with collocated subcarriers ($\eta=1$) to the BS for coarse delay estimation, which has the full unambiguous range $\frac{1}{\triangle f}$.
Based on which, the decimation factor $\eta^*$ is optimized and the sensing UE transmits signals with distributed subcarriers ($\eta^*>1$), which renders higher delay resolution, to refine the estimation results while avoiding the delay ambiguity.


\vspace{-0.2cm}
\subsection{Two-Stage Delay Estimation}\label{signal processing}
The signal processing procedures are summarized in {Algorithm} 2 and elaborated as follows.

\subsubsection{Stage-1}
\begin{figure} 
  \centering
  \includegraphics[width=0.38\textwidth]{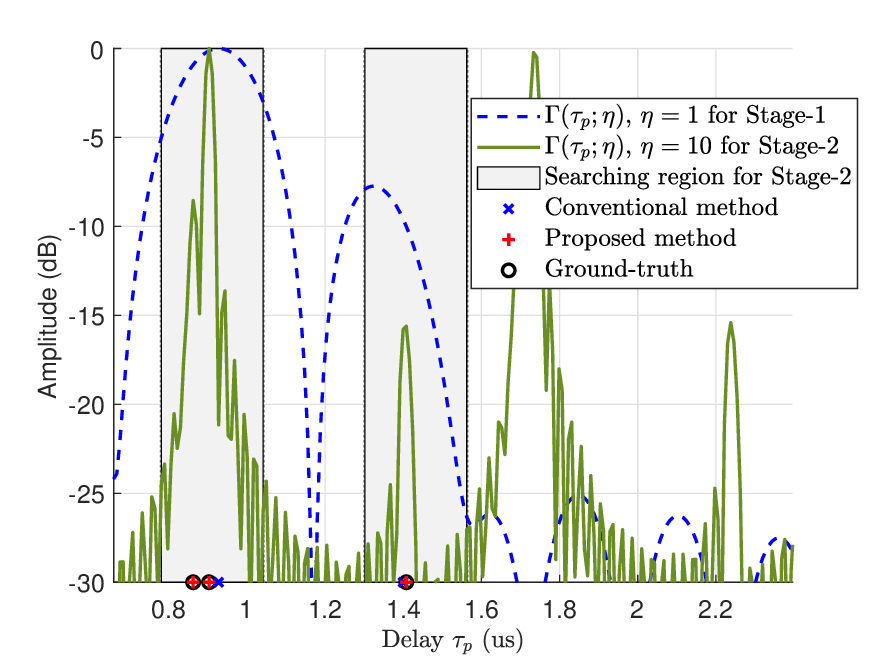}
  \caption{Comparison of the proposed TSDE method and the conventional method with collocated subcarriers.}\label{TSDS}\vspace{-10pt}
\end{figure}

The UE transmits uplink signals with collocated subcarriers for $\eta=1$ according to \eqref{DFT_spread}-\eqref{tx}, and then the BS processes the received signals and performs delay estimation based on Algorithm 1.
However, it suffers from poor delay resolution with $\tau_{\mathrm{res}}=\frac{1}{\triangle fK_1}$.
As shown in Fig.~\ref{TSDS}, when the delay difference of two paths is smaller than the delay resolution, such paths cannot be resolved.

However, the unambiguous range for $\eta=1$ is large, i.e., $\tau_{\mathrm{u}}=\frac{1}{\triangle f}$, which can be divided into $K_1={\tau_{\mathrm{u}}}/{\tau_{\mathrm{res}}}$ delay bins with each bin having the length of $\frac{\tau_{\mathrm{res}}}{T_s}=\frac{K}{K_1}=\eta_{\max}$.
The estimated delay indices $p_1,\cdots,p_{\hat{L}}$ are located into $u_1,\cdots,u_{L'}$ delay bins, where $L'\le \hat{L}$ in general due to the coarse resolution.
Denote by $\mathcal{U}\triangleq\{u_1,\cdots,u_{L'}\}$, which can be calculated as $\mathcal{U}=\mathrm{unique}(\{\lfloor \frac{p_l}{\eta_{\max}}\rfloor,p_l\in\mathcal{I}\})$, where $\mathrm{unique}(\mathcal{A})$ represents to select the unique elements of the set $\mathcal{A}$ and $\lfloor\cdot\rfloor$ denotes the integer flooring operation.
Based on this, a searching region $\Omega \triangleq \Omega_1\cup\cdots\cup\Omega_{L'}$ is defined for Stage-2, as shown in {\it Line 8} of Algorithm 2 and illustrated in the gray rectangles in Fig.~\ref{TSDS}.
Moreover, an optimal decimation factor $\eta^*>1$ will be selected for Stage-2, which aims to increase the delay resolution while avoid the grating lobes that may cause delay ambiguity to be located in the region $\Omega$.
The details for selecting $\eta^*$ will be discussed in Setion~\ref{performance}.

\subsubsection{Stage-2}
The UE transmits DFT-s-OFDM signal with distributed subcarriers for $\eta=\eta^*$ for refining the estimation results in Stage-1.
After the BS receives the signals, the same signal processing as in Algorithm 1 is performed but with $\eta=\eta^*$. Moreover, the delay index is searched in $\Omega$ with higher delay resolution $\tau_{\mathrm{res}}^*=\frac{1}{\triangle fK_1\eta^*}$.
As evident from Fig.~\ref{TSDS}, compared to the conventional one-stage method, the proposed method achieves higher delay resolution while avoiding the ambiguity without requiring additional bandwidth.

\subsection{Selection of $\eta^*$ and Performance Analysis}\label{performance}
\begin{figure} 
  \centering
  \includegraphics[width=0.38\textwidth]{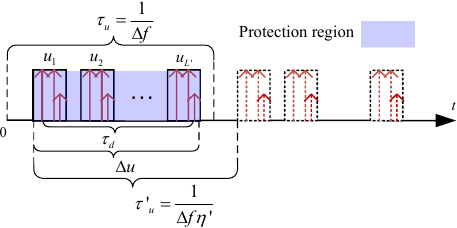}
  \caption{Illustration of the selection of $\eta^*$ for Stage-2 .}\label{MultiCluster}\vspace{-10pt}
\end{figure}

With the proposed method, after Stage-1, the multi-path delays may be located in different delay bins, within the region $\bar\Omega\triangleq \bar\Omega_1\cup\cdots\cup\bar\Omega_{L'}$, where $\bar\Omega_l=\Omega_lT_s=[u_l\tau_{\mathrm{res}},(u_l+1)\tau_{\mathrm{res}})$, $0\le u_l\le K_1-1$, $l=1,\cdots,L'$, as illustrated in Fig.~\ref{MultiCluster}.
Here, without loss of generality, we assume that $u_1<u_2<\cdots<u_{L'}$.
For $1\le\eta'\le \frac{K}{K_1}$, the delay distance to the nearest mirror delay bin where the grating lobe lies is $\tau_{\mathrm{u}}'=\frac{1}{\triangle f\eta'}$.
To avoid the ambiguity in Stage-2, we set a protection region that covers $\bar{\Omega}$ with the width $\triangle u=(u_{L'}+1-u_1)\tau_{\mathrm{res}}$, where $u_{L'}+1-u_1=\lceil\frac{\tau_d}{\tau_{\mathrm{res}}}\rceil\triangleq\xi$, with $\lceil\cdot\rceil$ being the integer ceiling operation.
For $0\le\tau_d<\tau_{\mathrm{u}}$, it has $1\le\xi\le K_1$ and $\triangle u=\xi\tau_{\mathrm{res}}\ge \tau_d$.
Note that if $\tau_{\mathrm{u}}'\ge\triangle u$, the grating lobe will not lie within the region $\Omega$, thus no ambiguity occurs, which requires that $1\le \eta'\le\frac{1}{\triangle f\triangle u }$.
With $1\le\eta'\le \frac{K}{K_1}$, thus the feasible $\eta'$ for Stage-2 without causing ambiguity should satisfy $1\le\eta'\le\min(\frac{1}{\triangle f\triangle u},\frac{K}{K_1})$.
Since the delay resolution increases with $\eta'$ as in \eqref{res}, the optimal $\eta^*$ is selected as
\begin{equation}\label{general}
\eta^*=\min(\frac{1}{\triangle f\triangle u},\frac{K}{K_1}).
\end{equation}
According to \eqref{general}, we derive the following theorem.

{\it Theorem 1}: For a DFT-s-OFDM based ISAC system with total bandwidth $B=\triangle f K$, the proposed TSDE method can achieve the full-bandwidth delay resolution $\frac{1}{B}$ and unambiguous range $\frac{K}{B}$, by using only $\frac{{B}\sqrt{\xi}}{\sqrt{K}}$ allocated bandwidth, as long as $\tau_d\le \frac{\sqrt{K\xi}}{B}$, for $1\le\xi\le \frac{K}{4}$.
\begin{IEEEproof}
According to \eqref{res}, to achieve the full-bandwidth delay resolution $\frac{1}{B}$, $\eta^*$ should be selected as $\frac{K}{K_1}$.
Based on \eqref{general}, to obtain $\eta^*=\frac{K}{K_1}$, it requires $\frac{1}{\triangle f\triangle u}\ge \frac{K}{K_1}$, which implies $K_1\ge B\triangle u$.
As $\triangle u=\xi\tau_{\mathrm{res}}$, we can derive that $K_1\ge\sqrt{K\xi}$.
Moreover, to guarantee the effectiveness of the propose method, $\eta^*$ should satisfy $\eta^*=\frac{K}{K_1}>1$.
As $\eta^*$ is an integer, it requires that $K_1\le \frac{K}{2}$.
Thus, we have $\sqrt{K\xi}\le K_1\le \frac{K}{2}$.
Therefore, to achieve the full-bandwidth delay resolution and unambiguous range, with the proposed method, the minimum bandwidth is $K_1\triangle f=\sqrt{K\xi}\triangle f=\frac{B\sqrt{\xi}}{\sqrt{K}}$ when $K_1=\sqrt{K\xi}$.
In this case, it requires that $\tau_d\le\triangle u=\xi\tau_{\mathrm{res}}$ and $\sqrt{K\xi}\le\frac{K}{2}$, where
$\tau_{\mathrm{res}}=\frac{1}{\triangle fK_1}=\frac{1}{\triangle f \sqrt{K\xi}}$.
Thus, we can derive that $\tau_d\le\frac{\sqrt{K\xi}}{B}$ and $1\le \xi\le \frac{K}{4}$.
\end{IEEEproof}
{\it Corollary 1.1}: When $\xi=1$, i.e., all the multi-path delays are located in a single delay bin after Stage-1, with $\tau_d\le\frac{\sqrt{K}}{B}=\tau_{\mathrm{res}}$, the proposed method can achieve the full-bandwidth delay resolution and unambiguous range, by using only $B/\sqrt{K}$ allocated bandwidth.

{\it Corollary 1.2}: When $\xi=\frac{K}{4}$, i.e., $\tau_d\le\frac{K}{2B}=\frac{\tau_{\mathrm{u}}}{2}$, the proposed method can achieve the full-bandwidth delay resolution and unambiguous range, by using only half-bandwidth $B/2$.

Note that the proposed TDSE method is effective for $0\le\tau_d\le\frac{\tau_{\mathrm{u}}}{2}$, which is suitable for most delay sensing scenarios, as it usually requires that $\tau_d<T_{\mathrm{cp}}$ to avoid the ISI while $\frac{\tau_{\mathrm{u}}}{2}=\frac{KT_s}{2}$ equals to the half of OFDM symbol duration, which is typically larger than the CP length $T_{\mathrm{cp}}$.

%

\begin{figure} 
    \centering
        \includegraphics[width=0.48\textwidth]{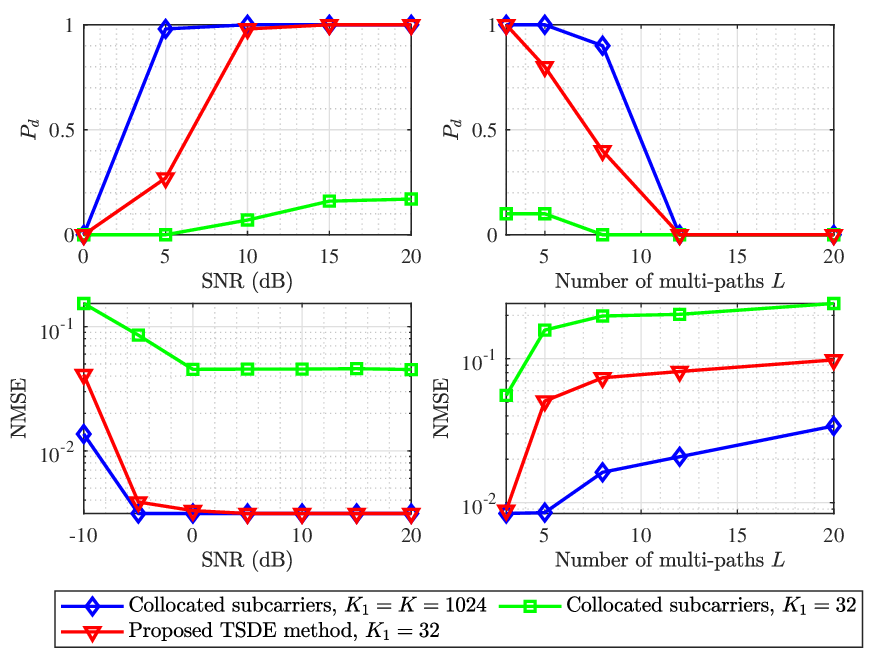}
    \caption{Compare the proposed TSDE method and conventional method in terms of $P_d$ and NMSE.}
    \label{compare}
    \vspace{-0.3cm} 
\end{figure}

\begin{figure} 
    \centering
        \includegraphics[width=0.38\textwidth]{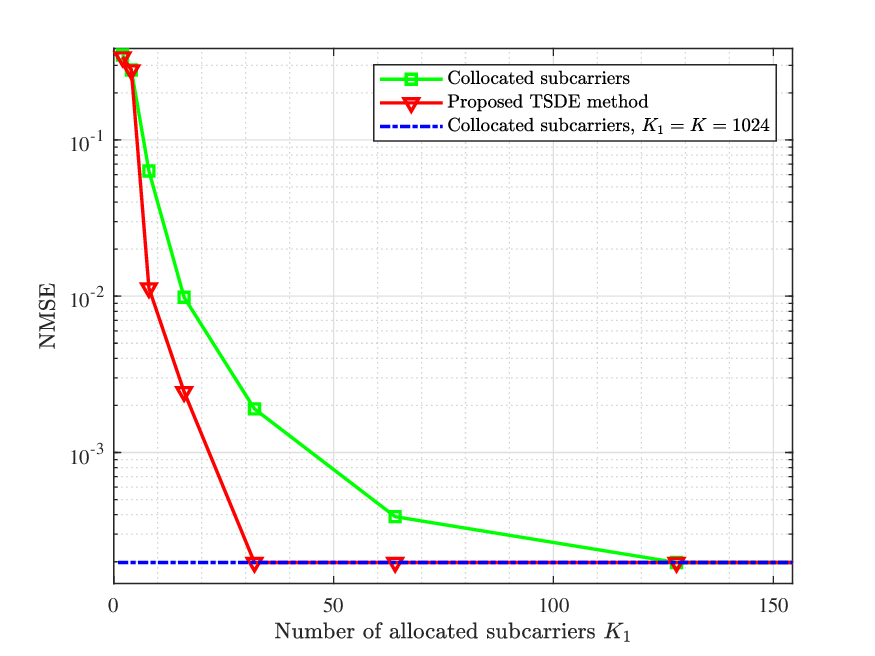}
    \caption{Compare the NMSE of the proposed TSDE method and conventional method versus the number of allocated subcarriers $K_1$.}
    \label{bandwidth}
    \vspace{-0.3cm} 
\end{figure}

\section{Simulation Results}
In this section, we evaluate the performance of the proposed TSDE method.
The subcarrier spacing is $\triangle f=120$ kHz, the total number of subcarriers is $K=1024$, and the full bandwidth of the system is $B=\triangle f K = 122.88$ MHz.
The number of subcarriers allocated to the sensing UE is $K_1=32$, which only occupies a partial bandwidth of $B_1=\triangle fK_1=3.84$ MHz.
The performance of the proposed TSDE method is compared to the conventional collocated subcarrier method for partial and full-bandwidth cases, respectively.

The probability of correct detecting the number of multi-paths, denoted by $P_d\triangleq\mathrm{Prob}\{\hat{L}=L\}$, and the normalized mean-square error (NMSE) of the delay estimations with the known of $L$, defined as $\mathrm{NMSE}=\mathbb{E}\big[\frac{1}{L}\sum\nolimits_{l=1}^{L}\frac{|\tau_l-\hat{\tau}_l|^2}{|\tau_l|^2}\big]$, are selected as the performance metrics.
From Fig.~\ref{compare}, it can be obtained that under the same bandwidth, the proposed TSDE method has higher $P_d$ and lower NMSE than the conventional method with collocated subcarriers.
This is expected, as the proposed TSDE method can achieve the higher delay resolution than the conventional method with the same bandwidth, as evident from Fig.~\ref{TSDS}.
Moreover, with suifficent high SNR (e.g., 10 dB), the proposed TSDE method performs comparably to the full-bandwidth case for $K_1=K=1024$.

Fig.~\ref{bandwidth} compares the NMSE of the proposed TSDE method with the conventional collocated subcarrier method as a function of the number of allocated subcarriers.
It is observed that as the allocated bandwidth (i.e., the number of allocated subcarriers $K_1$) increases, the proposed TSDE method can achieve the performance comparable to the conventional collocated subcarrier method with full-bandwidth, while utilizing only a fraction of the bandwidth.
This result validates the effectiveness of the proposed TSDE method.

%
%

\section{Conclusion}
In this letter, we proposed a new TSDE method for uplink DFT-s-OFDM based ISAC systems, which can achieve the full-bandwidth delay resolution and unambiguous range by only using a partial bandwidth.
Numerical results demonstrated that the proposed method achieves higher probability of correct detection and lower estimation NMSE compared with the conventional method with collocated subcarriers.

\bibliographystyle{IEEEtran}
\bibliography{sparseOFDM}

\end{document}